\pdfoutput = 1
\documentclass[prd,nofootinbib,preprint,preprintnumbers]{revtex4-2}
\usepackage{color}
\usepackage{amsmath,bm,amscd,mathrsfs}
\usepackage{environ,graphicx}
\usepackage{hyperref} 

\newcommand{\as}{\alpha_{\mathrm{s}}}

\newcommand{\LA}{\mathrm{A}}
\newcommand{\LB}{\mathrm{B}}
\newcommand{\scB}{\textsc{b}}

\newcommand{\LF}{\mathrm{F}}

\newcommand{\LJ}{\mathrm{J}}

\newcommand{\scR}{\textsc{r}}

\newcommand{\scT}{\mathrm{T}}
\newcommand{\LT}{\mathrm{T}}

\newcommand{\Lc}{\mathrm{c}}
\newcommand{\Ld}{\mathrm{d}}
\newcommand{\Le}{\mathrm{e}}

\newcommand{\Lg}{\mathrm{g}}

\newcommand{\Lp}{\mathrm{p}}

\newcommand{\Ls}{\mathrm{s}}

\newcommand{\Lu}{\mathrm{u}}

\newcommand{\LZ}{\mathrm{Z}}

\newcommand{\MSbar}{\overline{\mathrm{MS}}}

\newcommand{\GeV}{\ \mathrm{GeV}}
\newcommand{\TeV}{\ \mathrm{TeV}}

\newcommand{\mur}{\mu_{\textsc r}}

\newcommand{\muf}{\mu_{\textsc F}}

\newcommand{\cO}{{\cal O}}

\newcommand{\cT}{{\cal T}}

\definecolor{red}{rgb}{1,0,0}

\def\mi{{\mathrm i}}

\def\ket#1{\big|{#1}\big\rangle}
\def\bra#1{\big\langle{#1}\big|}
\def\<>#1{\big\langle{#1}\big\rangle}
\def\[]#1{\big[{#1}\big]}

\def\iket#1{|{#1}\rangle}

\newbox\charbox
\newbox\slabox
\def\s#1{{      
        \setbox\charbox=\hbox{$#1$}
        \setbox\slabox=\hbox{$/$}
        \dimen\charbox=\ht\slabox
        \advance\dimen\charbox by -\dp\slabox
        \advance\dimen\charbox by -\ht\charbox
        \advance\dimen\charbox by \dp\charbox
        \divide\dimen\charbox by 2
        \raise-\dimen\charbox\hbox to \wd\charbox{\hss/\hss}
        \llap{$#1$}
}}

\makeatother

\begin{document}

\title{From the quark parton model to QCD}

\author{Davison E.\ Soper}

\affiliation{
Institute for Fundamental Science\\
University of Oregon\\
Eugene, OR  97403-5203, USA
}

\email{soper@uoregon.edu}

\begin{abstract}
The quark parton model grew out of deeply inelastic scattering experiments. The parton model developed into a full theory, quantum chromodynamics, QCD. This article explains some of the physics issues encountered in connecting the parton model and QCD.
\end{abstract}

\date{25 June 2026}

\maketitle


\section{Deeply inelastic scattering}
\label{sec:DIS}

In 1966, the linear accelerator at what was then called the Stanford Linear Accelerator Center (SLAC) began operation. The first results were reported in Physical Review Letters in 1969 \cite{DIS1969A, DIS1969B}. The experiment was a very big undertaking. The theoretical interpretation of the results had, in many ways, a revolutionary impact on our understanding of fundamental physics. It is this interpretation that is the subject of this chapter. The description is presented as a logical development starting with the simplest ideas and leading to the properties of quantum chromodynamics (QCD), interwoven with some attention to the historical development of the subject.

\begin{figure}[t]
\begin{center}
\includegraphics[width = 6 cm]{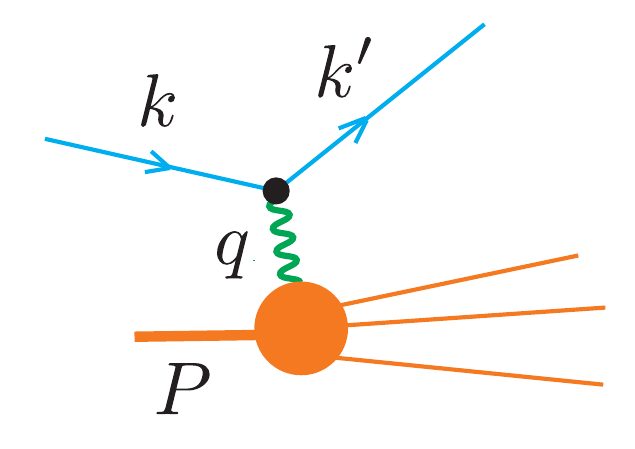}
\end{center}
\caption{
Amplitude for $\Le + \Lp \to \Le + X$. Only the outgoing electron is detected.}
\label{fig:DISdef}
\end{figure}

To understand the interpretation of the SLAC experiment, it is best to begin with some definitions. The experiment is electron-proton scattering in which only the scattered electron in the final state is measured. This is written as $\Le + \Lp \to \Le + X$, where $X$ stands for any final state particles, which are not measured. This process is depicted in Fig.~\ref{fig:DISdef}. The initial electron momentum is denoted by $k$ and final electron momentum is denoted by $k'$. The electron mass is small enough that we can treat the electron as massless, so that $k^2 = k^{\prime\,2} = 0$. The initial proton momentum is denoted by $P$. Letting $m_\Lp$ be the proton mass, we have $P^2 = m_\Lp^2$. The initial electron and proton have spin 1/2, but they are unpolarized. The spin of the final electron is not observed. The electron interacts with the proton by exchanging a virtual photon with momentum $q = k - k'$, as depicted in Fig.~\ref{fig:DISdef}.\footnote{It is also possible to exchange a Z boson, but at the energy available in the SLAC experiment, Z-boson exchange can be neglected.} Two kinematic variables are important for understanding the results:
\begin{equation}
\begin{split}
Q^2 ={}& -q^2
\;,
\\
x_\mathrm{bj} ={}& \frac{Q^2}{2 P\cdot Q}
\;.
\end{split}
\end{equation}
One uses $-q^2$ because $q^2$ is necessarily negative, The subscript on $x_\mathrm{bj}$ distinguishes this variable from other variables named $x$. It stands for James Bjorken, whose work on this subject we will describe below.\footnote{Bjorken was universally known by the name posted on a sign on his office door, ``bj'' with lower case letters. }

A third kinematic variable is
\begin{equation}
W^2 = (P+q)^2 = m_\Lp^2 + \frac{1-x_\mathrm{bj}}{x_\mathrm{bj}}\,Q^2
\;.
\end{equation}
Since $W^2$ must be greater than $m_\Lp^2$, we conclude that $x_\mathrm{bj} \le  1$. When $x_\mathrm{bj} < 1$, the scattering is inelastic. For small $Q^2$, resonant states like the $\Delta$ baryon can be produced, resulting in peaks in the cross section. For large $Q^2$ and $x_\mathrm{bj} < 1$, resonance peaks were not expected (or seen). One then says that we are exploring deeply inelastic scattering (DIS).

To describe the SLAC results, we work to first order in quantum electrodynamics (QED), so that the electron-photon vertex is given simply by $\mi e\gamma^\mu$. Then the cross section can be written as
\begin{equation}
d \sigma = \frac{4 \alpha^2}{s} \frac{d^3 {\vec k}^\prime}{2|{\vec k}^\prime|}
\frac{1}{(q^2)^2}\,
L^{\mu\nu}(k,q)\,W_{\mu\nu}(P,q)
\;.
\end{equation}
Here
\begin{equation}
L^{\mu\nu} =  \frac{1}{2}\,
{\rm Tr}\left(k\cdot \gamma\ \gamma^\mu k^\prime\cdot \gamma\ \gamma^\nu 
\right)
\;.
\end{equation}
In 1969, one could say very little about the tensor $W_{\mu\nu}(P,q)$, but one could say something. One knew that the photon coupled to the electromagnetic current, $J^\mu(x)$. With this information, one could define $W_{\mu\nu}$ using the state vector $\iket{P,s}$ for a proton with momentum $P$ and spin $s$:
\begin{equation}
\label{eq:WmunuJsJ0}
W_{\mu\nu}(P,q) = \frac{1}{4\pi}  \int\!d^4x\,e^{\mi q\cdot x}\,
\frac{1}{2}\sum_s \bra{P,s}J_\mu(x) J_\nu(0)\ket{P,s}
\;.
\end{equation}
One also knew that the current must be conserved: $\partial\cdot J(x) = 0$. This implies that $q^\mu W_{\mu\nu}(P,q) = 0$. Additionally, $W_{\mu\nu}(P,q)$ had to be real, symmetric under $\mu\leftrightarrow \nu$, and transform as a tensor under Lorentz transformations, including the parity transformation. These requirements imply that $W_{\mu\nu}(P,q)$ has a certain structure:
\begin{equation}
\begin{split}
\label{eq:Wmunuexpansion}
{{W_{\mu\nu}}} ={}&
-\left(
g_{\mu\nu}- \frac{q_\mu q_\nu}{q^2}
\right)
 {{F_1(x_\mathrm{bj},Q^2)}}
\\ &
+ 
\left(
P_\mu - q_\mu \frac{P\cdot q}{q^2}
\right)
\left(
P_\nu - q_\nu \frac{P\cdot q}{q^2}
\right)
\frac{1}{P \cdot q}\
 F_2(x_\mathrm{bj},Q^2)
\;.
\end{split}
\end{equation}
The functions $F_1$ and $F_2$ are called the proton structure functions. They are invariant under Lorentz transformations, so they must be functions of the only two scalar variables that one can form from $P$ and $q$, namely $Q^2$ and $x_\mathrm{bj}$. The notation in Eq.~(\ref{eq:Wmunuexpansion}) is the modern version of the notation used in 1969.

The SLAC experiment could measure $F_1$ and $F_2$, but there were no available predictions for them prior to the experiment. It was known that hadrons -- protons, neutrons, pions, {\em etc} -- interacted strongly with one another through what was called the strong interactions. However, there was no theory of the strong interactions.

One could try to use quantum field theory, as described in the textbook of the time by Bjorken and Drell \cite{BjorkenDrell}. This approach was successful for quantum electrodynamics.  For the strong interactions, one could perhaps use a quantum field for each hadron \cite {BjorkenDrell}. For quantum electrodynamics, there was a small coupling constant so that a perturbative expansion in powers of the coupling was useful. It was not useful for the strong interactions: whatever the fields and interactions were, the coupling constant was surely large.  

There was a non-relativistic quark model \cite{GellMannQuarks, ZweigQuarks}, in which hadrons were made of combinations of {\em up}, {\em down}, and {\em strange} quarks and antiquarks. A proton, for instance, is made of two up quarks and a down quark. Generally the particles, called baryons, that were similar to protons and neutrons, were made of three quarks. Their antiparticles were made of three antiquarks. Other particles, called mesons, were made of a quark and an antiquark. This model had some success in organizing the structure of the masses of hadrons and some of the properties of individual hadrons. However, it did not explain how hadrons interacted with one another or, indeed, how the quarks interacted inside of a hadron. This model did not have predictions for the SLAC experiments.

There was also S-matrix theory, as exemplified by the text of Goldberger and Watson \cite{GoldbergerWatson}. In this approach, one examined the scattering matrix that describes the scattering of hadrons in relativistic quantum mechanics. The S-matrix is an analytic function of the kinematic variables that characterize the scattering and it is a unitary matrix. With these and other general properties, one could say quite a lot about hadron scattering. One could hope that with a few more insights one could say much more. However, there were no predictions for the behavior of the SLAC structure functions.

There was one qualitative prediction that seemed sensible from any of these points of view. It was known that the proton has a finite size of order $1/m_\Lp$, or more exactly about $3/m_\Lp$. There was no known feature of protons or of the strong interactions that had a size scale smaller than this. Thus one could expect that for $Q^2 >  m_\Lp^2$, the structure functions would fall quickly as $Q^2$ increased. In this respect, they would be similar to the electromagnetic form factors of the proton.

There was a contrary prediction. In Eq.~(\ref{eq:WmunuJsJ0}), $W_{\mu\nu}(P,q)$ is given as a matrix element of a product of two current operators. In 1969, Bjorken analyzed the algebra of products of current operators \cite{BjorkenScaling} using an analysis similar to that used by Curtis Callan and David Gross to derive what is known as the Callan-Gross sum rule \cite{CallanGross}. Bjorken argued that it was plausible that $F_1(x_\mathrm{bj},Q^2)$ and $F_2(x_\mathrm{bj},Q^2)$ would be constant as $Q^2 \to \infty$ at fixed $x_\mathrm{bj}$. This property is called Bjorken scaling.

The first results of the experiment reported the behavior of $F_2(x_\mathrm{bj},Q^2)$, given an assumption for the ratio $F_2/F_1$ \cite{DIS1969A, DIS1969B}. The structure function $F_2(x_\mathrm{bj},Q^2)$ did not fall quickly with increasing $Q^2$. Instead, within the experimental errors, the behavior of $F_2(x_\mathrm{bj},Q^2)$ was consistent within Bjorken scaling.

\section{The parton model}
\label{sec:PartonModel}

Bjorken scaling could be explained by the current algebra argument of Ref.~\cite{BjorkenScaling}. However, a more appealing physical picture for the scaling results appeared around 1969: the parton model \cite{FeynmanPartons, DrellLevyYan, BjorkenPaschos}. We imagine that the proton is made of constituent particles that Feynman called partons. The description of the parton model in this chapter follows Ref.~\cite{BjorkenPaschos}.

The important properties of the partons are that some of them carry electric charge and that they are pointlike, like electrons. That is, they are not made of anything else and have no intrinsic size. The partons have interactions that hold them together to make a proton, so that a proton does have a size on the order of $1/m_\Lp$. We can imagine that, in a proton at rest, each parton has an interaction with another parton once in each time interval $\Delta t \sim 1/m_\Lp$. The parton model does not specify what the partons are or what their interactions with each other are.

\begin{figure}[t]
\begin{center}
\includegraphics[width = 6 cm]{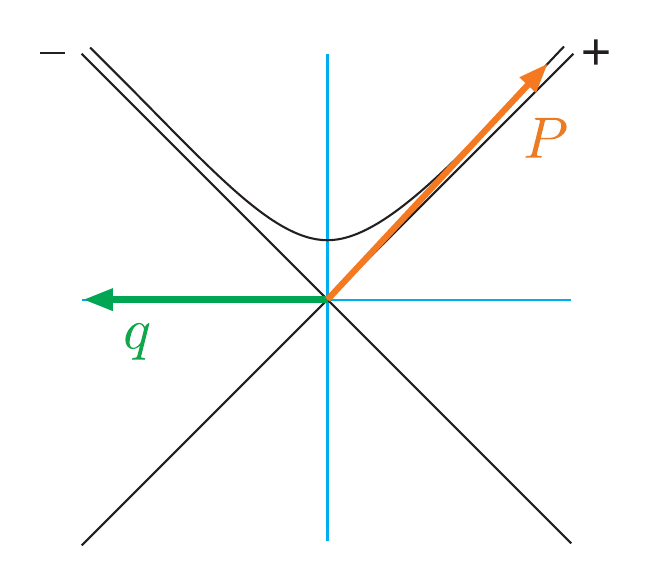}
\end{center}
\caption{
The Breit frame. The proton momentum $P$ and the virtual photon momentum $q$ have zero transverse components.}
\label{fig:BreitFrame}
\end{figure}

We choose to study DIS in a particular reference frame, the Breit frame, in which the momentum transfer $q$ lies along the negative $z$ axis and the proton has positive $P^3$ and zero transverse momentum, as illustrated in Fig.~\ref{fig:BreitFrame}

In order to facilitate the use of this reference frame, it is helpful to use what are called null-plane components for all Lorentz vectors. If $v$ is any vector with components $(v^0, v^1, v^2, v^3)$ in the customary description, we use an alternative description with components $(v^+, v^-, v^1, v^2)$, with
\begin{equation}
v^\pm = \frac{1}{\sqrt 2}\,(v^0 \pm v^3)
\;.
\end{equation}
The dot product between any two vectors is then
\begin{equation}
a\cdot b = a^+ b^- + a^- b^+ - \bm a \cdot \bm b
\;.
\end{equation}
We use boldface letters for the transverse components $(a^1, a^2)$ of a vector, with
\begin{equation}
\bm a \cdot \bm b = a^1 b^1 + a^2 b^2
\;.
\end{equation}
In particular
\begin{equation}
v^2 = 2 v^+ v^- - \bm v^2
\;.
\end{equation}

To help develop an intuition for the dynamics of partons, it is useful to note that fully relativistic quantum mechanics analyzed using null-plane coordinates has a symmetry that is the same as the symmetry of nonrelativistic quantum mechanics in two space dimensions. In this correspondence, the transverse components of the momentum of a quantum state, $\bm p$, are considered as a momentum in two dimensions. Then $p^+$ plays the role of the nonrelativistic mass of the state.  Finally, $p^-$ plays the role of the Hamiltonian. For any two dimensional vector $\bm v$, the system is invariant under \cite{KogutSoper}
\begin{equation}
\begin{split}
p^+ \to{}& p^+
\;,
\\
\bm p \to{}& \bm p + p^+ \bm v
\;,
\\
p^- \to{}& p^- + \bm p \cdot \bm v + \frac{1}{2}\,p^+ \bm v^2
\;.
\end{split}
\end{equation}
On one hand, this is the form of a nonrelativistic boost in two dimensions. On the other hand, it is a Lorentz transformation: $p^2 = 2p^+ p^- - \bm p^2$ is invariant. This correspondence helps us to appreciate the form of $p^-$ for an on-shell parton:
\begin{equation}
p^- = \frac{\bm p^2}{2 p^+}
\;.
\end{equation}

The dot product of $p$ with a space-time position vector $x$ is
\begin{equation}
p\cdot x = p^- x^+ + p^+ x^- - \bm p \cdot \bm x
\;.
\end{equation}
The coordinate that is Fourier conjugate to the null-plane hamiltonian $p^-$ is $x^+$. Thus we can understand $x^+$ as playing the role of time for the evolution of partons. Indeed, one can usefully quantize quantum field theory on planes of equal $x^+$ instead of planes of equal $x^0$ \cite{Dirac,KogutSoper}. We will not need to develop this viewpoint in depth in this chapter. 

In the Breit frame, the momentum transfer has components
\begin{equation}
(q^+,q^-, \bm q) = \frac{1}{\sqrt{2}}\,(- Q, Q, \bm 0)
\;.
\end{equation}
The incoming proton has momentum components
\begin{equation}
(P^+, P^-, \bm P) \approx 
\frac{1}{\sqrt{2}}\left(\frac{Q}{x_{\mathrm bj}}, \frac{x_{\mathrm bj}}{Q}\,m_\Lp^2, \bm 0\right)
\;.
\end{equation}
Here $P^2 = m_\Lp^2$ exactly, but for large $Q$, $P^+$ is larger than $P^-$ by a factor $Q^2/(m_\Lp^2 x_\mathrm{bj}^2)$. For this reason, we have approximated $2 P\cdot q$ by $2 P^+ q^-$ in calculating $x_{\mathrm bj}$ from $P$ and $q$. We learn from this that in our chosen reference frame $P^+$ is very large when $Q$ is large.

Evidently, the proton momentum has undergone a large Lorentz boost compared to what it was in the proton rest frame. The notation using null-plane components has the useful property that under a Lorentz boost  in the $z$ direction with hyperbolic angle $\omega$, vectors transform according to $v \to \bar v$ with
\begin{equation}
\begin{split}
\bar v^+  ={}& e^\omega v^+
\;,
\\
\bar v^-  ={}& e^{-\omega} v^-
\;,
\\
\bar {\bm v} ={}& \bm v
\;.
\end{split}
\end{equation}
This is much simpler than the representation of the same boost using the customary description, in which $v^0$ and $v^3$ mix with coefficients $\cosh \omega$ and $\sinh \omega$. The proton momentum in the proton rest frame was $P_\mathrm{rest}^+ = m_\Lp/\sqrt 2$. Thus the boost angle is given by
\begin{equation}
e^\omega = \frac{P^+}{P_\mathrm{rest}^+}
\approx \frac{Q}{m_\Lp x_{\mathrm bj}}
\;.
\end{equation}

This brings us to the key observation of the parton model. In the Breit frame, the interactions among the partons are slowed down by a large factor $e^\omega$. If the null-plane time $\Delta x^+$ between interactions was of order $\Delta x^+ \sim 1/m_\Lp$ in the proton rest frame, then in the Breit frame it is of order 
\begin{equation}
\Delta \bar x^+ \sim e^\omega \frac{1}{m_\Lp} \approx \frac{Q}{m_\Lp^2 x_{\mathrm bj}}
\;.
\end{equation}
The null-plane time for the interaction with the virtual photon can be considered to be
\begin{equation}
\Delta \bar x_\gamma^+ \sim \frac{1}{q^-} = \frac{\sqrt 2}{Q}
\;.
\end{equation}
Thus when $Q$ is large, the null-plane time for the interaction with the virtual photon is very short, while the null-plane time between interactions among the partons is very long. That is, for the purpose of considering the scattering of a parton by the virtual photon, we can consider the parton to be effectively a free, noninteracting particle.

\begin{figure}[t]
\begin{center}
\includegraphics[width = 6 cm]{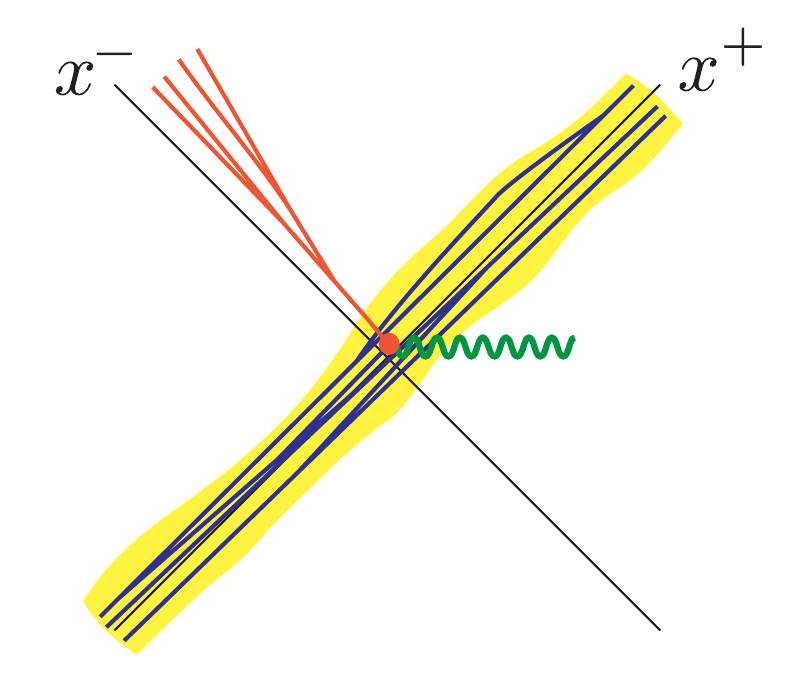}
\end{center}
\caption{
The parton model for deeply inelastic scattering in $x^+$-$x^-$ coordinates. The lines represent the paths of the partons.}
\label{fig:partonspacetime}
\end{figure}

This picture is illustrated in Fig.~\ref{fig:partonspacetime}. The partons in the proton move approximately parallel to the $x^+$-axis. They interact, but only slowly as measured in $x^+$. At a null-plane time that we can take to be $x^+ \approx 0$, one of the partons is scattered by the virtual photon. It then moves in the $x^-$ direction. It can still have interactions, but now its interactions are slow as measured in $x^-$.

The parton interactions determine the probability to find a parton with a given momentum. There could be several types of partons. We label the types with an index $i$ and let $Q_i$ be the charge of a parton of type $i$ in units of the proton charge. Let $f_{i/\Lp}(\xi)\,d\xi$ be the probability to find a parton of type $i$ in a proton with a fraction $\xi$ (within a range $d\xi$) of the +-momentum of the proton: $p^+ = \xi P^+$. The parton also has a transverse momentum $\bm p$, but $f_{i/\Lp}(\xi)$ is the probability integrated over $\bm p$. We assume that $\bm p$ is small enough compared to $Q$ that we can use the approximation $\bm p \approx \bm 0$ when we consider the scattering by the virtual photon. We also assume that the parton mass and the proton mass are small enough compared to $Q$ that we can approximate them by zero.

After the scattering, the parton momentum is $\xi P + q$. The scattered parton should be on shell:
\begin{equation}
0 = (\xi P + q)^2 = 2\xi P\cdot Q - Q^2
\;.
\end{equation}
This gives us $\xi = Q^2/(2 P\cdot q)$:
\begin{equation}
\xi = x_\mathrm{bj}
\;.
\end{equation}
We can now use can use lowest order quantum electrodynamics to calculate the scattering cross section. This calculation, using spin 1/2 for the charged partons, gives
\begin{equation}
\label{eq:DISpartonmodel}
F_2(x_\mathrm{bj}, Q^2) = \sum_i Q^2_i\, x_\mathrm{bj}\,f_{i/\Lp}(x_\mathrm{bj})
\;.
\end{equation}
Remarkably, the parton model gives Bjorken scaling.

The functions $f_{i/\Lp}(\xi)$ are called parton distribution functions (PDFs). It is important to distinguish the parton distribution functions from the DIS structure functions $F_1(x_\mathrm{bj}, Q^2)$ and $F_2(x_\mathrm{bj}, Q^2)$. The structure functions are given directly from experiment. They are related to parton distribution functions, but this relation becomes more complicated than Eq.~(\ref{eq:DISpartonmodel}) when we go beyond the simple parton model.

The parton model gives us Bjorken scaling, but it does not tell us what kinds of partons there are or what their charges $Q_i$ are. It does not specify the interactions of the partons, so there is no possibility of it giving us the parton distribution functions.

It was quickly realized that the parton model can give predictions for other kinds of experiment \cite{DrellLevyYan1, DrellLevyYan2, DrellLevyYan3}. 

Consider first a collision of two hadrons, A and B, with $s = (p_\LA + p_\LB)^2 \gg m_\Lp^2$. Then in the center-of-momentum frame both hadrons have a very large energy $E \gg m_\Lp$. Thus we can view both incoming hadrons as being made of partons, with the interactions among the partons slowed by relativistic time dilation so that they are approximately free on a time scale $1/\sqrt s$. A parton of type $i$ in hadron A can annihilate with a parton that is an antiparticle of this type, with charge $-Q_i$, in hadron B. This produces a virtual photon that can split into a muon and an antimuon. Thus there must be a process $A + B \to \mu^+ + \mu^- + X$ with a cross section proportional to
\begin{equation}
\sum_i f_{i/A}(\xi_\LA)\, f_{\bar i/B}(\xi_{\LB})\, Q_i^2
\;.
\end{equation}
Letting the squared dimuon mass be $Q^2$, we have $\xi_\LA \xi_\LB = Q^2/s$. The remaining factors in the formula for the cross section then follow using lowest order quantum electrodynamics, given a hypothesis for the spin of the charged partons. This is the Drell-Yan process \cite{DrellYan}. The formula is not fully predictive unless we know the types $i$ of partons and the parton distribution function for each type. However, the formula for the Drell-Yan cross section does make a prediction about how the cross section scales as $s$  and $Q^2$ increase with fixed $Q^2/s$. This process was in fact observed in 1970 in proton-Uranium collisions \cite{DrellYanSeen}. Muon pairs with $Q^2$ up to about $20\GeV^2$ were observed, qualitatively consistent with the parton model result.

We can also consider the total cross section to make hadrons in the collision of a high energy electron with a high energy positron: $\Le^+ + \Le^- \to \mathrm{hadrons}$ \cite{DrellLevyYan3}. Whatever partons are, they must be pointlike elementary particles and some of them must carry charge. An electron-positron collision can create a virtual photon that can split into any elementary particle and its antiparticle. If we suppose that the charged partons have spin 1/2, then the cross section to make a parton and antiparton of type $i$ should be $Q_i^2$ times the cross section to make a $\mu^+\mu^-$ pair. This implies that the ratio $R$ of the cross section for $\Le^+ + \Le^- \to \mathrm{hadrons}$ to the cross section for $\Le^+ + \Le^- \to \mu^+ + \mu^-$ should be
\begin{equation}
\label{eq:Rratio}
R = \sum_i Q_i^2
\;.
\end{equation}
This experiment was carried out at SLAC starting in 1972, with results consistent with Eq.~(\ref{eq:Rratio}). We will have more to say about the SLAC $e^+e^-$ experiment in Sec.~\ref{sec:Charmonium}.

\section{Nature and interactions of the partons}
\label{sec:PartonIdentity}

Without specifying what the partons were and how they interacted, the parton model was of limited usefulness. The successful proposal was that the charged partons were quarks and antiquarks. At first impression, this idea could not be exactly right because the quarks of the quark model were what can be called {\em constituent quarks}. There were just three of them in a proton and they were clearly not pointlike. One can best think of a constituent quark as a parton quark surrounded by a cloud of other partons. 

There is another feature of the constituent quarks that at first seems puzzling but then offers a significant insight. The constituent quarks carry what we now call flavor: up, down, or strange. They carry spin 1/2, and they carry momentum. In the quark model, the wave functions of protons and other baryons are symmetric under interchange of their flavor, spin, and momentum quantum numbers. However, the quarks are fermions, so their wave functions should be antisymmetric under interchange of all of their quantum numbers. This can be fixed by assigning to each quark another quantum number $i$, called {\em color}, that takes three values $i = 1,2,3$. Then the wave function can be antisymmetric in the color quantum numbers of the quarks \cite{GreenbergQuarks, HanNambu}. 

If the colors of the three quarks in a proton are $i,j,k$ then the wave function can be the antisymmetric tensor $\epsilon_{ijk}$ (with $\epsilon_{123} = 1$, $\epsilon_{213} = - 1$, {\em etc.}), This gives the model a symmetry: if $\psi_i(x)$ is a quark field with color $i$, then the theory is invariant under $\psi_i(x) \to \sum_j U_{ij} \psi_j(x)$ where $U_{ij}$ is a unitary matrix with determinant 1. The product of any two such matrices gives a unitary matrix with determinant 1, so these matrices form a group, known as SU(3). A matrix in SU(3) can be written in the form
\begin{equation}
U = \exp\Big(\mi \sum_{a = 1}^8 \theta_a t^a \Big)
\;,
\end{equation}
where the $\theta_a$ are eight parameters and there are eight matrices $t^a$, which are $3\times 3$ Hermitian matrices with trace 0. These are the generator matrices of SU(3). The corresponding invariance property of the theory is often called {\em global} SU(3) invariance.

This suggests an important insight. Why not consider a quantum field theory with quark fields $\psi_i(x)$ that is invariant under SU(3) transformations that can depend on the space-time coordinate $x$ \cite{FritzschGellMannLeutwyler}? That is
\begin{equation}
\label{eq:gaugetransformation}
U(x) = \exp\Big(\mi \sum_{a = 1}^8 \theta_a(x)\, t^a \Big)
\;.
\end{equation}
This invariance is called SU(3) {\em gauge invariance}. It requires that we introduce into the theory eight vector fields $A_\mu^a(x)$. This gives an SU(3) {\em gauge theory}.

The theory with just a simple global invariance under the group U(1), with phase factors $U = \exp(\mi\theta)$ can be made into a U(1) gauge theory by incorporating just one field $A_\mu(x)$. This theory is quantum electrodynamics The field $A_\mu(x)$ is the electromagnetic potential. When we extend the theory based on the global SU(3) group of quark color to a gauge group, the theory is quantum chromodynamics.

A warning is needed at this point. In order to derive Feynman rules for either U(1) gauge theory or SU(3) gauge theory, one has to break the gauge invariance of the theory by introducing a non-gauge-invariant factor into the theory. This is called choosing a gauge. There are many ways to do this. For instance, one can choose Feynman gauge. If one is calculating something that is gauge invariant, like a cross section, then one is guaranteed to get the same result independent of the choice of gauge. The choice of gauge is beyond our scope here.

\begin{figure}[t]
\begin{center}
\includegraphics[width = 6 cm]{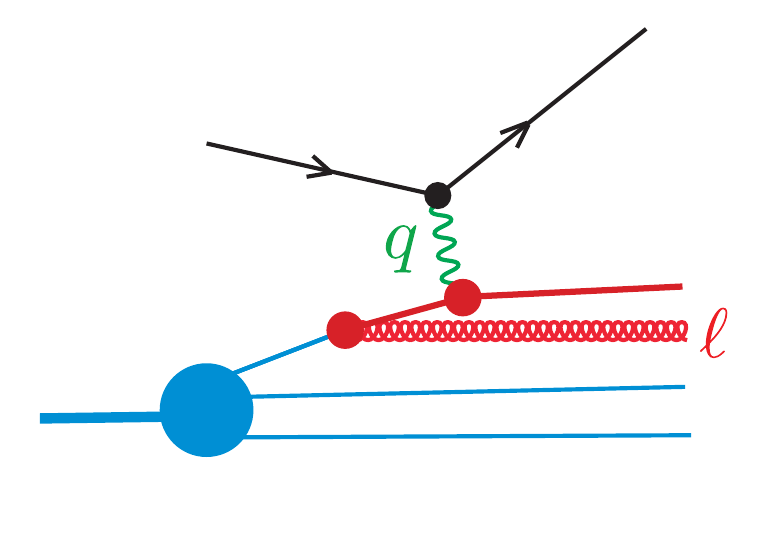}
\end{center}
\caption{
Deeply inelastic scattering with a gluon emission.}
\label{fig:DISNLOgraph}
\end{figure}

Turning the parton model into QCD is a big step forward, but it creates a problem. Consider a contribution to DIS depicted in Fig.~\ref{fig:DISNLOgraph}. A quark in a proton is scattered by a virtual photon with momentum $q$, but, before the scattering, the quark emits a gluon with momentum $\ell$. It is certainly possible for this interaction to happen long before (as measured by $x^+$) the scattering by the virtual photon. This possibility, which corresponds to $\ell^-$ being small, is consistent with the parton model. The gluon is a parton and the emission of the gluon simply corresponds to an interaction among the partons in the proton. However, the gluon emission can happen at almost the same $x^+$ as the scattering. This corresponds to large $\ell^-$. When one works this out by evaluating the Feynman diagram, there is an integration over $\ell^-$ with an effective upper cutoff $\ell^- \lesssim Q$ and a lower cutoff coming from the quark wave function $\ell^- \gtrsim m_\Lp$. There is now a contribution proportional to $\as \log(Q^2/m_\Lp^2)$, where 
\begin{equation}
\as = \frac{g^2}{4\pi}
\end{equation}
is the QCD analog of the electromagnetic fine structure constant $\alpha = e^2/(4\pi)$ and $g$ is the QCD coupling. Thus Bjorken scaling is broken.

One can claim that the breaking of Bjorken scaling is not so bad because $\log(Q^2/m_\Lp^2)$ is a slowly varying function of $Q^2$. This small violation of Bjorken scaling is not really ruled out by the SLAC experimental results. However, if one gluon emission leads to a small problem, what about 20 gluons being emitted, with a contribution $\as^{20} \log^{20}(Q^2/m_\Lp^2)$? It seems clear that QCD cannot be consistent with the observed experimental results unless $\as$ in these calculations is small. This does not seem plausible because $\as$ is the coupling of the strong interactions. It should be large, not small.

\section{The running coupling}
\label{sec:RunningCoupling}

\begin{figure}[t]
\begin{center}
\includegraphics[width = 5 cm]{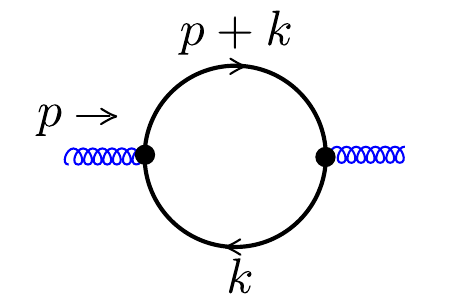}
\end{center}
\caption{
Feynman graph for the gluon self-energy.}
\label{fig:gluonselfenergy}
\end{figure}

At this point, there is something that we need to understand about the coupling in a quantum field theory that has a dimensionless coupling, like $e$ for quantum electrodynamics or $g$ for QCD. Feynman diagrams like those shown in Fig.~\ref{fig:gluonselfenergy} contain an integration over the momentum $k$ that runs around the loop. The integration over $k$ is ultraviolet divergent: there is a divergence coming from the region of large $k$. Large $k$ corresponds to short space-time distances between the points corresponding to the vertices of the diagram. Now, we know nothing about the true behavior of the contributing diagrams at distances much smaller than $1/(100 \TeV)$, say. Presumably some new theory takes over, but we do not know what that new theory might be. The best that we can do is to impose a cutoff: we can pick a scale parameter $\mur$ and remove the part of the integration that comes from $k > \mur$. Here we can mean that that absolute value of each component of $k$ is required to be smaller than $\mur$. This gives what is known as the {\em renormalized} value of the graph.

We note that most graphs with integrations over a loop momentum $k$ do not get significant contributions from very large $k$. That is, they are convergent in the ultraviolet. We simply compute these graphs with no cutoff.

The cutoff described above is very crude. A much better version is to let $\epsilon$ be a small positive parameter and to perform the integration in $4 - 2 \epsilon$ dimensions. This results in a pole $1/\epsilon$. We subtract the pole plus a specified extra part. The remainder with $\epsilon \to 0$ is the renormalized value of the graph according to what is called the $\MSbar$ renormalization prescription. This program can be extended to higher order graphs.

The result of the renormalization program applied to QCD is to produce a version of QCD that depends on $\mur$. Analysis of the renormalization procedure leads to a conclusion that is perhaps surprising \cite{CallanRenGroup,SymanzikRenGroup}. Consider two versions of QCD defined by two different values of $\mur$. Then both define the same physical theory but with different values of the coupling: $\as(\mu_{\scR,1})$ and $\as(\mu_{\scR,2})$. The quark masses also depend on $\mur$, but the masses of the up, down, and strange quarks are small enough that they can be set to zero for many purposes. The normalizations of the quantum fields depends on $\mur$, but these normalizations do not affect cross sections. Thus the most important fact is that the coupling depends on $\mur$. The dependence is given by a differential equation:
\begin{equation}
\label{eq:asevolution}
\mur\,\frac{d\as(\mur)}{d \mur}
= 8\pi\beta\!\left(\frac{\as(\mur)}{4\pi}\right)
\;.
\end{equation}
The function $\beta$ has a perturbative expansion in powers of $\as$ starting with a term proportional to $\as^2$.

To understand what this means, consider the cross section for $\Le^+ + \Le^- \to \mathrm{hadrons}$ as a function of $Q^2$, where $Q$ is the momentum of the virtual photon formed by the annihilation of the colliding electron and positron. When calculated using QCD combined with lowest order electrodynamics, the cross section has the form
\begin{equation}
\begin{split}
\label{eq:sigmaee}
\sigma_{\mathrm tot} ={}& \frac{4\pi \as}{Q^2}
\left(\sum_a Q_a^2\right)\!
\left\{
1 + \frac{\as(\mur)}{\pi }
+ \left[
1.4092 + 1.9167\log\!\left(\frac{\mur^2}{Q^2}\right)
\right]
\left(\frac{\as(\mur)}{\pi}\right)^2
+ \cO(\as^3)
\right\}
\;.
\end{split}
\end{equation}
Here there is a factor of the fine structure constant $\alpha$ and a sum over the flavors $a$ of quarks. The first QCD correction is proportional to $\as(\mur)$. The next term has an explicit $\mur$ dependence, which is just what is needed to cancel the $\mur$ dependence of $\as(\mur)/\pi$. If we truncate the series at this point, then the residual $\mur$ dependence of the cross section is of the order of the first omitted term, $\as^3$. Thus, up to the order calculated, the physical results are independent of $\mur$ even though the coupling $\as$ depends on $\mur$.

There is another lesson to be learned from Eq.~(\ref{eq:sigmaee}). Nominally, we have complete freedom to choose the value of the renormalization scale $\mur$. However, this freedom is not so complete. The second term has a contribution $\log(\mur^2/Q^2)$. If we were to choose $\mur^2$ to be very different from $Q^2$, this logarithm would be large and could make the second term as large as the first. Then the usefulness of perturbation theory would be spoiled. Thus we should choose $\mur^2 = Q^2$ or, more generally, $\mur^2/Q^2$ of order 1. This is despite the intuition that if we think of $\mur$ as the value of an ultraviolet cutoff, it should be much larger than $Q^2$.

We can generalize this lesson: when applying QCD to a problem with a physical scale $Q^2$, we should choose the renormalization scale close to $\mur^2 = Q^2$.

\section{Asymptotic freedom}
\label{sec:AsymptoticFreedom}

We can now return to the question encountered at the end of Sec.~\ref{sec:PartonIdentity}: if the strong interactions are strong, how can it be that when we look at a proton with deeply inelastic scattering, the interactions of the partons seem to be weak. The surprising answer is that when one calculates the leading term in the $\beta$ function in Eq.~(\ref{eq:asevolution}), this leading term is negative \cite{GrossWilczek, Politzer}. Thus even though $\as$ is large when $\mur$ is smaller than 1 GeV, $\as(\mur)$ becomes smaller as $\mur$ increases. At $\mur = 2 \GeV$, $\as \sim 0.3$. When $\mur$ is the mass of the Z boson, 91 GeV, $\as(M_\LZ) \approx 0.118$. This phenomenon is called {\em asymptotic freedom} because quarks and gluons act as almost free, noninteracting, particles at short distance scales.

This happens in gauge theories like SU(3). In quantum electrodynamics, with the gauge group U(1), $\alpha(\mur)$ grows as $\mur$ increases. Perturbation theory is useful for quantum electrodynamics with large momenta not because $\alpha(\mur)$ decreases as $\mur$ grows but because $\alpha(\mur)$ is small over a wide range of scales.

\section{Confinement}
\label{sec:Confinement}

Interactions with small momentum transfers among partons are described by Feynman graphs with $\as(\mur)$ with small $\mur$. In this case $\as(\mur)$ is large. 

We know from experiment that there are hadrons in nature: protons, neutrons, rho mesons, pions, {\em etc}. We understand these to be bound states of partons. QCD perturbation theory is not useful for calculating their properties, but we can use lattice gauge theory for this purpose.

We also know from experiment that there are no free quarks or gluons. This phenomenon is called {\em confinement}: quarks and gluons are always confined inside hadrons.

Confinement can be understood using the following picture, known as the {\em Lund string model} \cite{LundString}. Start with a quark and an antiquark with opposite colors. Pull them apart. Then a string structure forms in which color electric field between them is contained in a tube. This color string carries a substantial amount of energy per unit length. If there are no light quarks in the theory, the potential energy of quark and antiquark color sources that are fixed in space is proportional to the distance between them. This phenomenon is seen in lattice QCD \cite{LatticePotential}. Of course, quarks and antiquarks produced in a high energy collision are not fixed in space but are moving away from each other, subject to the force produced by the string. Furthermore the string has its own dynamics because it carries an energy density and a string tension. 

In QCD with light quarks, when the quark and antiquark produced by a high energy interaction have separated enough, it is energetically favorable to use the energy contained in a length of string by letting the string break, creating a new quark and new antiquark. After more string breakings, one can obtain several quark-antiquark pairs, which form several mesons. More complicated processes can also produce baryons.

This picture produces the model of hadron production from partons in the parton shower event generator \textsc{Pythia} \cite{Pythia}.

\section{The November revolution}
\label{sec:Charmonium}

In 1974, the basics of QCD as the theory of the strong interactions were in place. However, quarks and gluons were an abstraction. They were parts of calculations, but they did not seem to be parts of the real world. That changed on November 11, 1974. An experiment at Brookhaven National Laboratory measured the momenta of $\Le^+ \Le^-$ pairs produced in collisions of protons with beryllium. The cross section was plotted as a function of the mass $W$ of the $\Le^+ \Le^-$ pair, $W^2 = (p_{\Le^-} + p_{\Le^+})^2$. This plot showed a sharp peak near $W = 3.1 \GeV$ \cite{BNLcharm}. In an experiment at SLAC, electrons were made to collide with positrons with an adjustable $\Le^+ \Le^-$ mass $W$, $W^2 = (p_{\Le^-} + p_{\Le^+})^2$. By varying $W$ in small steps, the cross section for $\Le^+ + \Le^- \to \mathrm{hadrons}$ could be plotted as a function of $W$. This plot showed a sharp peak at $W = 3.105 \GeV$ \cite{SLACcharm}. Both results were announced at SLAC on the same day. Within a day, a substantial fraction of the particle physicists around the world heard of the result by telephone conversations with each other. (There was no email.)

The result had a simple interpretation. A new kind of quark, the charm quark $\Lc$, had been produced together with its antiquark $\bar\Lc$. The charm quark was heavy, so that $c\bar c$ pairs could not be produced in low energy collisions. The existence of charm quarks had been hypothesized in order to make the theory of weak interactions work better, but hadrons containing charm quarks had not been seen. A $\Lc$ quark and $\bar \Lc$ antiquark with opposite colors could form a bound state, similar to other $q\bar q$ mesons, but with a mass that turned out to be 3.105 GeV. This bound state could decay back to light quarks by making an off-shell photon, but the lifetime for this decay would be long, so that the peak in the cross section for making the bound state followed by its decay would be narrow compared to $3.1\GeV$. This explanation fit the facts nicely. The new $\Lc\bar\Lc$ meson is now called the $\LJ/\psi$.

The discovery of the $\LJ/\psi$ changed QCD by adding a new quark $\Lc$ with a mass of 1.3 GeV. Beyond the new quark, QCD theory remained unchanged. However, after this discovery, quarks in general seemed to be not just an abstraction but an observable part of nature. Quarks and gluons were to be taken very seriously.

As noted earlier in Eq.~(\ref{eq:Rratio}), the ratio $R$ of the cross section for $\Le^+ + \Le^- \to \mathrm{hadrons}$ to the cross section for $\Le^+ + \Le^- \to \mu^+ + \mu^-$ should be
\begin{equation}
\label{eq:Rratiobis}
R = 3 \sum_a Q_a^2
\;.
\end{equation}
Here we sum over the flavors $a$ of quarks, $\Lu, \Ld, \Ls$ as long as the $\Le^+ \Le^-$ mass $W$ is less than than than 3.1 GeV. Then we multiply by 3 for the three colors of each kind of quark. Thus for $W < 3.1 \GeV$,
\begin{equation}
\label{eq:Rratiolow}
R_\mathrm{low} = 3 \left[\left(\frac{2}{3}\right)^2 + \left(-\frac{1}{3}\right)^2
+ \left(-\frac{1}{3}\right)^2\right]
= 2
\;.
\end{equation}

For $\Le^+ \Le^-$ mass $m$ substantially greater than than 3.1 GeV (but less than 9 GeV, where yet another heavy quark appears), we should sum over $\Lu, \Ld, \Ls, \Lc$. The result is then modified by an easily derived function $f(m_\Lc/W)$ that accounts for the mass of the charm quark:
\begin{equation}
\label{eq:Rratiohigh}
R_\mathrm{high} = 3 \left[\left(\frac{2}{3}\right)^2 + \left(-\frac{1}{3}\right)^2
+ \left(-\frac{1}{3}\right)^2 + \left(\frac{2}{3}\right)^2
f\!\left(\frac{m_\Lc}{W}\right)
\right]
= \frac{10}{3} + \frac{4}{3}\left[f\!\left(\frac{m_\Lc}{W}\right) - 1\right]
\;.
\end{equation}
The experimental results from SLAC were in reasonable agreement with these predictions. The experimental results would not be even close to matching the theory without the factor 3 for the number of colors.

\section{Structure of the final state in QCD}
\label{sec:Splittings}

Because of asymptotic freedom, perturbation theory is relevant in QCD for processes that involve a large momentum transfer. Let us consider, then, $\Le^+ + \Le^- \to \mathrm{hadrons}$ when $Q^2 = (p_{\Le^+} + p_{\Le^-})^2$ is very large. We use Feynman diagrams for partons: quarks, antiquarks, and gluons at a finite order of perturbation theory. What does the final state look like? A definitive answer requires serious calculations and analyses. However, a quick look can provide intuition.

At order $\as$, one can produce three partons, a quark, an antiquark, and a gluon. Let $E_3$ be the energy of the gluon and let $\theta_{13}$ be the angle between the gluon and the quark. Calculation gives
\begin{equation}
\label{eq:eetoqqg}
\frac{1}{\sigma_0}\,
\frac{d\sigma}{dE_3\,d\cos \theta_{13}}
=
\frac{\alpha_{\rm s}}{2\pi}\,C_{\rm F}\,
\frac{f(E_3,\theta_{13})}
{E_3(1- \cos\theta_{13})}
\;.
\end{equation}
Here $\sigma_0$ is the cross section for $\Le^+ + \Le^- \to q + \bar q$ at order 0 in $\as$ and $C_\LF = 4/3$ is a factor related to the color in the emission of a gluon from a quark or antiquark. The function $f(E_3,\theta_{13})$ is straightforward and nonsingular when $E_3 \to 0$ or $\theta_{13} \to 0$, but slightly complicated. We see that the cross section becomes large when the gluon is ``soft,'' $E_3 \to 0$ and also when the gluon and the quark become collinear, $\theta_{13} \to 0$. In fact, if we were to integrate this cross section over $E_3$ with a small lower cutoff $E_3^\mathrm{min}$, the result would be proportional to $\log(Q/E_3^\mathrm{min})$. If we were to integrate this cross section over $\cos(\theta_{13})$ with a small lower cutoff on $1 - \cos(\theta_{13})$, the result would be proportional to the logarithm of the cutoff. We should not rely on this calculation for very small $E_3$ or $1 - \cos(\theta_{13})$ because in this regime we should use higher orders of perturbation theory. But we can expect that the splitting of a quark to a quark plus a gluon is most probable when the daughter partons are close to being collinear or the daughter gluon is soft.

\begin{figure}[t]
\begin{center}
\includegraphics[width = 6 cm]{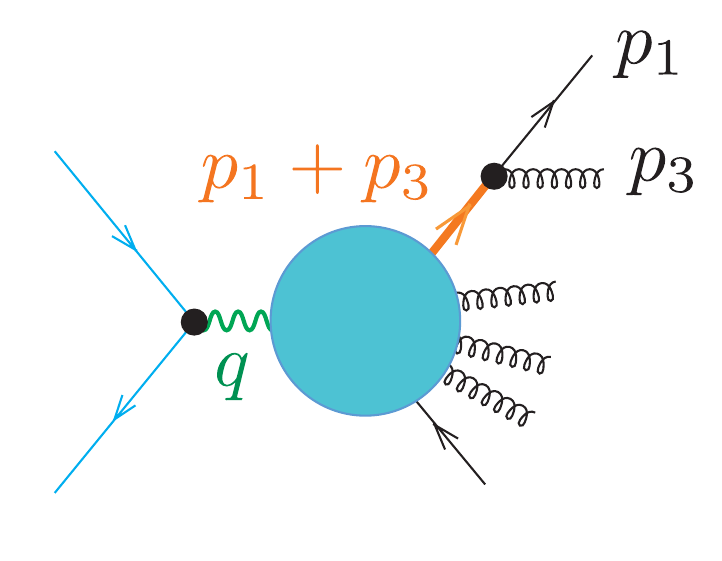}
\end{center}
\caption{
Illustration of infrared singularities in the amplitude for $\Le^- + \Le^+ \to \mathrm{hadrons}$.}
\label{fig:eetojets}
\end{figure}

This situation is generic in QCD. Suppose that there are several partons in the final state, say $m+1$ of them, produced by a sum of high order Feynman diagrams. Let one of them be a quark with momentum $p_1$ and let one of them be a gluon with momentum $p_3$ as depicted in Fig.~\ref{fig:eetojets}. Then there is a diagram in which these two partons are produced by the splitting of a quark with momentum $p_1 + p_3$. With the daughter partons on shell, $p_1^2 = p_3^3 = 0$, we have
\begin{equation}
(p_1 + p_3)^2 = 2 p_1\cdot p_3
= 2 E_1 E_3 (1-\cos\theta_{13})
\;.
\end{equation}
The Feynman rules dictate that the Feynman amplitude is proportional to a Dirac spinor factor and a factor 
\begin{equation}
\frac{1}{(p_1 + p_3)^2} = \frac{1}{2 E_1 E_3 (1-\cos\theta_{13})}
\;.
\end{equation}
When we square this to get a cross section and account for the Dirac spinor factor and the integration measure for integrating over the momenta of the final state partons, the result is a generalization of Eq.~(\ref{eq:eetoqqg}): a cross section that is singular for $E_3 \to 0$ and for $\theta_{13} \to 0$, with
\begin{equation}
d\sigma \propto \frac{dE_3\,d\cos\theta_{13}}{E_3 (1-\cos\theta_{13})}
\;.
\end{equation}

Thus there should be a large contribution to the cross section when the quark and the gluon are close to being collinear or the gluon is close to being soft. There are similar large contributions to the cross section when two gluons are close to being collinear or one of them is close to being soft. These large contributions come from a splitting $\Lg \to \Lg + \Lg$. A splitting $\Lg \to q + \bar q$ gives a large contribution when the quark and antiquark are close to being collinear, but not when one of them is close to being soft.

This argument leads to an expectation for the distribution of partons obtained after a perturbative calculation in QCD for $\Le^+ \Le^-$ annihilation at very large $Q^2$. We expect that the partons will be grouped into a few sprays of partons, which are called {\em jets}. Each jet consists of one or several partons with momenta almost collinear with each other. In addition, there should be some partons with small momenta in directions at large angles to the jets. Cross sections to make jets are discussed in Sec.~\ref{sec:jets}

With confinement, one sees hadrons, not partons, in experiments. However, the formation of hadrons involves small momentum transfers among partons. Small momentum transfers should not much change the directions or the energies of the parton jets. Thus we expect to see jets of hadrons, with some additional hadrons with small momenta in directions at large angles to the jets.

\section{Infrared safety}
\label{sec:InfraredSafety}

We have seen that asymptotic freedom for QCD gives us the possibility of calculating the total cross section for $\Le^+ + \Le^- \to \mathrm{hadrons}$ at large $Q^2 = (p_{\Le^+} + p_{\Le^-})^2$. We calculate the cross section as a power series expansion in $\as(Q^2)$, which is useful because $\as(Q^2)$ is small when $Q^2$ is large. We can also calculate cross sections in which a measurement is applied to the final state hadrons.

We should be a bit careful, because we cannot usefully calculate just anything concerning the final state hadrons produced in such a collision. A simple example of something in $\Le^+ + \Le^- \to \mathrm{hadrons}$ that one can measure and compare to a  QCD calculation is the {\em thrust} distribution. If there are $m$ hadrons with momenta $\{p_1,\dots, p_m\}$ produced a particular event, the thrust for that event is
\begin{equation}
\label{eq:thrust}
\cT(\{p_1,\dots,p_m\}) = \max_{\vec u} 
\frac{\sum_{i=1}^m |\vec p_i \cdot \vec u |}{\sum_{i=1}^m |\vec p_i| }
\;.
\end{equation}
Here $\vec u$ is a unit vector, $|\vec u| = 1$. We choose $\vec u$ to maximize the indicated ratio.

Suppose that some of the hadrons have momenta closely aligned with a vector $\vec u_\mathrm{J}$ and the others have momenta closely aligned with a vector in the opposite direction.  We can imagine that an event like this occurs when the electron and positron produce a quark and an antiquark and that the quark turns into a narrow jet of hadrons moving in one direction, while the antiquark turns into  a narrow jet of hadrons moving in the opposite direction. Then the total momentum of the partons aligned with $\vec u_\mathrm{J}$ is approximately $(Q/2)\,\vec u_\mathrm{J}$ and the total momentum of the partons anti-aligned with $\vec u_\mathrm{J}$ is approximately $-(Q/2)\,\vec u_\mathrm{J}$. In this case, the optimum vector is $\vec u \approx \vec u_\mathrm{J}$ and the thrust is close to 1. Not all events will look like this. Events with less alignment will have $\cT < 1$.

Given this definition of thrust for a given event, we can define a thrust observable $F$ by
\begin{equation}
\label{eq:FforThrust}
F(\{p_1,\dots,p_m\}) =
\theta(T_\LA < \cT(\{p_1,\dots,p_m\}) < T_\LB)
\;.
\end{equation}
When applied to an event with $m$ hadrons, $F$ measures whether its thrust is between $T_\LA$ and $T_\LB$. We can measure the cross section for the thrust to lie between $T_\LA$ and $T_\LB$ using
\begin{equation}
\begin{split}
\label{eq:sigmaFee}
\sigma[{F}] ={}& 
\frac{1}{2!} \int d \Omega_2\
\frac{d \sigma[2]}{d \Omega_2}\
F(\{p_1,p_2\})
\\
& +
\frac{1}{3!} \int d \Omega_2 d E_3 d \Omega_3\
\frac{d \sigma[3]}{d \Omega_2 d E_3 d \Omega_3}\
F(\{p_1,p_2,p_3\})
\\
& +
\frac{1}{4!} \int d \Omega_2 d E_3 d \Omega_3 d E_4 d \Omega_4\
\frac{d \sigma[4]}{d \Omega_2 d E_3 d \Omega_3 d E_4 d \Omega_4}\
F(\{p_1,p_2,p_3,p_4\})
\\
&
+ \cdots .
\end{split}
\end{equation}

As noted at the start of this section, we cannot hope to reliably calculate just anything about the distribution of hadrons in $\Le^+ \Le^-$ annihilation at very large $Q^2$. Is this $F$ something that we could reliably calculate? Let us generalize this question and consider a set of functions $F(\{p_1,\dots,p_m\})$ defined for $m$ particle momenta such that $F$ is symmetric under interchange of any two of the momenta. Let us calculate using not hadron momenta but parton momenta. Based on what we learned about the structure of final states in QCD perturbation theory, we can claim that $F$ can be reliably calculated if it is {\em infrared safe} in the following sense.

Pick any number $m+1$ of partons and any constant $z$ with either $z = 0$ or $0 < z < 1$. Then $F$ is infrared safe if, for any such choice,
\begin{equation}
\begin{split}
\label{eq:IRsafety}
F(\{p_1,\dots,(1-z)p_\LJ,z p_\LJ\}) ={}& 
F(\{p_1,\dots,p_\LJ\})
\;.
\end{split}
\end{equation}
This says that if two of the $m+1$ partons become collinear or if one of them becomes infinitely soft, $p\to 0$, then $F$ for that configuration matches $F$ for just $m$ partons, with $p_m = (1-z)p_\LJ + z p_\LJ$. In this situation, the cross section does not change. (It does not matter which partons become collinear or soft because $F$ is a symmetric function of the parton momenta.)

In a perturbative calculation, there will be an infrared singularity in the limit of a collinear or soft emission. However, there are also infrared singular virtual loop graphs, in which nothing is emitted. The unitarity of the S matrix dictates that the sum of the probability to emit the parton plus the probability not to emit the parton is 1. Thus the cross section if finite as long as the measurement $F$ does not distinguish whether the parton was emitted or not. We conclude that when the observable $F$ is infrared safe, we can calculate the cross section $\sigma[F]$ in perturbation theory and get a finite answer.

We can conclude a little more by examining what happens when $\{p_1,\dots,p_m, p_{m+1}\}$ is close to the singular limit $\{p_1,\dots,(1-z)p_\LJ,z p_\LJ\}$, but not exactly at the limit. We can say that $F$ is infrared safe at scale $\mu_F$ if $F(\{p_1,\dots,p_m, p_{m+1}\})$ is a good approximation to $F(\{p_1,\dots,p_\LJ\})$ when $(p_m - (1-z)p_\LJ)^2 < \mu_F^2$ and $(p_{m+1} - z p_\LJ)^2 < \mu_F^2$. That is, parton splittings below the scale $\mu_F$ do not matter for $F$. If the infrared safety scale $\mu_F$ is large compared to the $1\GeV$ scale of parton binding in a hadron, then it should be a good approximation to use the parton-level calculation and apply it to the measured cross section defined with hadron momenta. A simple way to test whether $\mu_F$ for the observable $F$ is large enough is to calculate $\sigma[F]$ using a parton shower event generator that includes a model for hadronization and observe whether turning hadronization on and off makes a substantial difference for $\sigma[F]$.

This definition of infrared safety was given in general. The simplest example of an infrared safe observable in $\Le^+ \Le^-$ annihilation is the total cross section, for which
\begin{equation}
F(\{p_1,\dots,p_m\}) = 1
\;.
\end{equation}
The observable $F$ that measures thrust using Eqs.~(\ref{eq:thrust}) and (\ref{eq:FforThrust}) is more complicated. With a little analysis, one verifies that it is infrared safe.

The idea of infrared safe observables outlined here is logically a step in developing full QCD from the parton model. It was not a part of arguments presented in the 1970s, but was incorporated into QCD reasoning in the following decade.

\section{The QCD improved parton model}
\label{sec:DISPDFs}

Let $d\sigma/d^3 \vec k^\prime$ be the cross section for DIS in which $k'$ is the momentum of the scattered electron. Then the parton model result (\ref{eq:DISpartonmodel}) extended to full QCD is
\begin{equation}
\begin{split}
\label{eq:DISfactorization}
\frac{d \sigma}{d^3 \vec k'}
={}&
\int_0^1\! d \xi \sum_a\, f_{a/\Lp}(\xi,\muf)\,
\frac{d \hat\sigma_{a}(\xi,\muf)}{d^3 \vec k'}
+{\cal O}(m_\Lp/Q)
\;.
\end{split}
\end{equation}
Here we do not use the decomposition into structure functions $F_1$ and $F_2$, but we work at lowest order in quantum electrodynamics, so that the validity of the decomposition into structure functions is guaranteed. On the right-hand side of Eq.~(\ref{eq:DISfactorization}), $f_{a/\Lp}(\xi,\muf)$ is the parton distribution function that represents the probability per unit $\xi$ to find a parton of type $a$ in a proton, carrying a fraction $\xi$ of the proton momentum. Here $a$ labels the flavor of parton: $\Lg,\Lu, \bar{\Lu},\Ld, \bar{\Ld},\dots$. The label $a$ now does not include the parton's color. Rather $f_{a/\Lp}(\xi,\muf)$ is the probability density to find a parton of flavor $a$ of {\em any} color. The parton distribution function depends on a {\em factorization scale} $\mu_\LF$ that we will discuss below. The factor $d \hat\sigma_{a}(\xi,\muf)/d^3 \vec k^\prime$ is a quantity that we are to calculate in perturbative QCD. At lowest order in perturbation theory, $d \hat\sigma_{a}(\xi,\muf)/d^3 \vec k^\prime$ is proportional to $\delta(\xi - x_\mathrm{bj})$, so that we recover Eq.~(\ref{eq:DISpartonmodel}).

In the DIS experiment at SLAC, nothing was measured about the hadronic final state in addition to measuring the momentum $k'$ of the scattered electron. One could generalize the simplest measurement by defining a more general observable based on a function $F(\vec k', \{p_1,\dots, p_m\})$, where $\{p_1,\dots, p_m\}$ are the momenta of final state hadrons. Then we could measure the cross section for $F$ by a simple generalization of Eq.~(\ref{eq:sigmaFee}) for $\Le^+ \Le^-$ annihilation:
\begin{equation}
\begin{split}
\label{eq:sigmaFDIS}
\sigma[{F}] ={}& 
\int\!d^3\vec k'
\frac{1}{2!} \int d \Omega_2\
\frac{d \sigma[2]}{d\vec k' d \Omega_2}\
F(\vec k',\{p_1,p_2\})
\\
& +
\int\!d^3\vec k'
\frac{1}{3!} \int d \Omega_2 d E_3 d \Omega_3\
\frac{d \sigma[3]}{d\vec k' d \Omega_2 d E_3 d \Omega_3}\
F(\vec k',\{p_1,p_2,p_3\})
\\
& +
\int\!d^3\vec k'
\frac{1}{4!} \int d \Omega_2 d E_3 d \Omega_3 d E_4 d \Omega_4\
\frac{d \sigma[4]}{d\vec k' d \Omega_2 d E_3 d \Omega_3 d E_4 d \Omega_4}\
F(\vec k',\{p_1,p_2,p_3,p_4\})
\\
&
+ \cdots .
\end{split}
\end{equation}
We suppose that $F$ be infrared safe in the sense of Eq.~(\ref{eq:IRsafety}). We also require
\begin{equation}
\label{eq:IRsafetyA}
F(\vec k',\{p_1,\dots,p_m,p_{m+1}\})
= F(\vec k',\{p_1,\dots,p_m\})
\end{equation}
when $p_{m+1}$ is collinear with the proton momentum $P$ (treating $P$ as massless). That is, final state hadrons with small transverse momenta in the Breit frame should not affect the measurement. As long as the observable $F$ is infrared safe in this sense, we can calculate $\sigma(F)$ in perturbative QCD using
\
\begin{equation}
\begin{split}
\label{eq:Afactorization}
\sigma[F] ={}& \sum_{a} \int_0^1\!d\xi\,
f_{a/A}(\xi, \muf)\,
\hat\sigma_{a}[F;\xi, \muf]
+ \cO(m_\Lp/Q)
\;.
\end{split}
\end{equation}

The DIS cross section (\ref{eq:DISfactorization}) and its generalization (\ref{eq:Afactorization}) are calculable in the perturbative QCD extension of the parton model. Nevertheless, we recognize that we are calculating with QCD partons and hadrons are not partons. Thus we are missing effects from interactions among the final state partons that involve momentum transfers below 1 GeV. For that reason, we have included an error term ``$+\cO(m_\Lp/Q)$'' that does not disappear no matter how many orders of perturbation theory we use.

Now a note of caution is in order. We found in Sec.~\ref{sec:PartonIdentity} that there is a problem with the parton model when it is applied to QCD. When we consider a QCD Feynman diagram in which a quark emits a gluon before the quark is scattered by the virtual photon, as in Fig.~\ref{fig:DISNLOgraph}, it appears that Bjorken scaling is broken. It is helpful to analyze this breaking in a little detail.

Let the incoming quark have momentum
\begin{equation}
p = (\xi P^+, 0, \bm 0)
\;,
\end{equation}
with $0 < \xi < 1$. Here we treat the quark as being massless and having just a small transverse momentum, which we neglect. Let the emitted gluon have transverse momentum $\bm \ell$ and a fraction $(1-z)$ of the + momentum of the initial state quark, with $0 < z < 1$. Thus the momentum components of the gluon momentum are
\begin{equation}
\ell = 
\left((1-z)\xi P^+, \frac{\bm \ell^2}{2(1-z)\xi P^+}, \bm \ell
\right)
\;.
\end{equation}
After the gluon emission, the quark has momentum
\begin{equation}
\tilde p = 
\left(z \xi P^+, \frac{\bm \ell^2}{2z\xi P^+}, -\bm \ell
\right)
\;.
\end{equation}

We integrate over $z$ and $\bm \ell$ inside the Feynman diagrams for the cross section and we denote the probability for the incoming quark to emit the gluon by $(\as/\pi)\, G(z,\bm \ell^2)/\bm \ell^2$. This probability is simply obtained from the Feynman diagram. Then the DIS cross section in this picture is
\begin{equation}
\frac{d\sigma}{d^3\vec k'} =
\sum_a
\int_0^1\! d \xi\, f_{a/\Lp}(\xi)\,
\int_0^1\!dz \int\!\frac{d^2\bm \ell}{\bm \ell^2}\,
\frac{\as}{2\pi^2}\, G(z,\bm \ell^2)\,
\frac{d \hat\sigma_{a,0}(z\xi,\bm \ell)}{d^3 \vec k'}
\;.
\end{equation}

We can estimate the null-plane time $\Delta x^+$ between the gluon emission and the scattering of the quark as the inverse of $\tilde p^-$:
\begin{equation}
\Delta x^+ = \frac{2z\xi P^+}{\bm \ell^2}
\;.
\end{equation}
Recall that $P^+ \approx Q/(\sqrt 2 x_\mathrm{bj})$, so that if $\bm\ell^2$ is of the order of $m_\Lp^2$, then $\Delta x^+$ is very large compared to $1/m_\Lp$. This is the parton model picture, in which the partons interact slowly. The momentum fraction of the quark as seen by the virtual photon is $z\xi$. However, when we examine the Feynman diagram, we find that we should integrate over $\bm\ell$ with a measure $d^2\bm\ell/\bm \ell^2$ up to roughly $\bm\ell^2 \sim Q^2$. This breaks the parton model picture.

It is easy to see what to do. We choose a parameter $\muf$ and break the integration over $\bm \ell$ into $0 < \bm \ell^2 < \muf^2$ and $\muf^2 < \bm\ell^2 < \infty$. The region $\muf^2 < \bm\ell^2 < \infty$ is counted as arising from an order $\as$ correction to $d \hat\sigma_{a}(\xi)/d^3 \vec k^\prime$. That is, there is a correction,
\begin{equation}
\Delta \frac{d \hat\sigma_a(\xi,\muf)}{d^3 \vec k'} =
\int_0^1\!dz \int\!\frac{d^2\bm \ell}{\bm \ell^2}\,
\theta(\bm \ell^2 > \muf^2)\,
\frac{\as}{2\pi^2}\, G(z,\bm \ell^2)\,
\frac{d \hat\sigma_{a,0}(z\xi,\bm \ell)}{d^3 \vec k'}
\;,
\end{equation}
to the lowest order quark scattering cross section.

This leaves the region $0 < \bm \ell^2 < \muf^2$, which is counted as arising from an order $\as$ change in the parton distribution function,
\begin{equation}
\label{eq:PDFchange}
\Delta f_{a/\Lp}(\xi')
= \int_0^1\! \frac{d\xi}{\xi}\, f_{a/\Lp}(\xi)\,
\int_0^1\!dz \int\!\frac{d^2\bm \ell}{\bm \ell^2}\,
\theta(\bm \ell^2 < \muf^2)\,
\frac{\as}{2\pi^2}\, G(z,\bm \ell^2)\,
\delta\!\left(z - \frac{\xi'}{\xi}\right)
\;.
\end{equation}
Then this part of the cross section is (after changing integration variable from $z$ to $\xi' = z \xi$)
\begin{equation}
\Delta\frac{d\sigma}{d^3\vec k'} =
\sum_a
\int_0^1\! d \xi'\, \Delta f_{a/\Lp}(\xi')\,
\frac{d \hat\sigma_{a,0}(\xi', \bm 0)}{d^3 \vec k'}
\;.
\end{equation}
Here we can neglect $\bm \ell$ in $d \hat\sigma_{a,0}(z\xi,\bm \ell)/d^3 \vec k'$.

The parameter $\muf$ is called the {\em factorization scale}. We can choose the value of $\muf$. Because of the cutoff at $\muf$, the parton distribution function now depends on $\muf$. Of course, this dependence breaks Bjorken scaling, but this breaking is rather mild. We will return to the $\muf$ dependence of the parton distribution functions in Sec.~\ref{sec:DGLAP}.

This same QCD improved parton model can be applied to high energy reactions with two incoming hadrons, A and B. The first proposed example was the Drell-Yan process, $A + B \to \mu^+ + \mu^- + X$, with large dimuon squared mass $Q^2$. Here we do not measure anything about the hadrons in the final state. More generally, one could measure an observable $F$ that depends on the momenta $\{p_1,\dots,p_m\}$ of final state hadrons, as in Sec.~\ref{sec:InfraredSafety}. The observable should be infrared safe in a more stringent sense than in Sec.~\ref{sec:InfraredSafety}. We require Eq.~(\ref{eq:IRsafety}), which covers the case that two hadrons become collinear to each other. We also require
\begin{equation}
\label{eq:IRsafetyAB}
F(\{p_1,\dots,p_m,p_{m+1}\})
= F(\{p_1,\dots,p_m\})
\end{equation}
when $p_{m+1}$ is collinear with $p_A$ or $p_B$ (treating $p_A$ and $p_B$ as massless). That is, in an experiment with colliding hadron beams with momenta along the $\pm z$ axis, final state hadrons with small transverse momenta should not affect the measurement. We can include the measurement of a $\mu^+$, $\mu^-$ pair, but here we should demand that the measurement not be sensitive to the transverse momentum of the pair for small transverse momentum.

For an observable in hadron-hadron collisions that is infrared safe in this sense, the cross section in the QCD improved parton model is
\begin{equation}
\begin{split}
\label{eq:ABfactorization}
\sigma[F] ={}& \sum_{a,b} \int_0^1\!d\xi_A \int_0^1\!d\xi_B\
f_{a/A}(\xi_A, \muf)\, f_{b/B}(\xi_B, \muf)\,
\hat\sigma_{ab}[F;\xi_A, \xi_B,\muf]
+ \cO(m_\Lp/Q)
\;.
\end{split}
\end{equation}
Here $\hat\sigma_{ab}[F;\xi_A, \xi_B,\muf]$ is a quantity that we are to calculate using quarks and gluons in perturbative QCD. Now there are two parton distribution functions: $f_{a/A}(\xi_A, \muf)$ represents the probability per unit $\xi_A$ to find a parton of type $a$ in hadron $A$ carrying a fraction $\xi_A$ of the hadron momentum and $f_{b/B}(\xi_B, \muf)$ represents the probability density to find a parton of type $b$ in hadron $B$ carrying a fraction $\xi_B$ of the hadron momentum. In Eq.~(\ref{eq:ABfactorization}), there are corrections of order $m_\Lp$ divided by a scale $Q$ characteristic of the hard scattering observable $F$.

Within a decade following the DIS experiment at SLAC, it was recognized that the parton model should be extended to include perturbative QCD calculations using something like Eqs.~(\ref{eq:DISfactorization}), (\ref{eq:Afactorization}), and (\ref{eq:ABfactorization}) \cite{AmatiPetronzioVeneziano, AmatiPetronzioVeneziano2, EllisGeorgietal, LibbySterman}.

These equations express a property of the cross sections known as {\em PDF factorization}. This factorization should work at any order in perturbation theory for $\hat \sigma$. This property is often called simply ``factorization'' but there are many properties of quantities related to perturbative QCD that carry this name. This property is also sometimes called ``collinear factorization,'' but this name is sometimes used with other meanings. Thus it seems useful to distinguish this factorization as PDF factorization. The factors here are $\hat \sigma$ and one or two parton distribution functions. It is not obvious that PDF factorization is valid for hadron-hadron collisions. We will return to this question in Sec.~\ref{sec:factorization}.

Early arguments \cite{AmatiPetronzioVeneziano, AmatiPetronzioVeneziano2, EllisGeorgietal, LibbySterman} for PDF factorization concerned particular cross sections rather than cross sections for general infrared safe observables as presented in this section. The $m_\Lp/Q$ power correction for DIS in Eq.~(\ref{eq:DISfactorization}) can be treated in more detail than presented here. The derivation for DIS uses something called the operator product expansion. Then  the $m_\Lp/Q$ power corrections are called {\em higher twist} contributions.

\section{PDF evolution}
\label{sec:DGLAP}

In the parton model adapted to QCD, the parton distribution functions $f_{a/A}(\xi,\muf)$ depend on the factorization scale $\muf$. The $\muf$ dependence is easy to understand if we use the example from Sec.~\ref{sec:DISPDFs} of a quark splitting within hadron $A$, $q \to q + \Lg$. Eq.~(\ref{eq:PDFchange}) gives the contribution to the quark PDF from this gluon emission graph. There is an integration over the transverse momentum $\bm \ell$ of the gluon over the range $|\bm\ell| < \muf$. Differentiating this equation with respect to $\muf$ gives
\begin{equation}
\begin{split}
\muf\frac{df_{a/\Lp}(\xi')}{d\muf}
={}& \int_{\xi'}^1\! \frac{d\xi}{\xi}\, f_{a/\Lp}(\xi)\,
\int\!\frac{d^2\bm \ell}{\bm \ell^2}\,
\muf\,\frac{d\theta(|\bm \ell| < \muf)}{d\muf}
\frac{\as}{2\pi^2}\, G(\xi'/\xi,\bm \ell^2)
\\
={}& \int_{\xi'}^1\! \frac{d\xi}{\xi}\, f_{a/\Lp}(\xi)\,
\int\!\frac{d^2\bm \ell}{|\bm \ell|}\,
\delta(|\bm \ell|- \muf)
\frac{\as}{2\pi^2}\, G(\xi'/\xi,\bm \ell^2)
\;.
\end{split}
\end{equation}
When $\muf^2$ is much smaller than $Q^2$, $G(\xi'/\xi,\bm \ell^2)$ is independent of $\bm \ell^2$. Then we obtain a simple equation for how the distribution of quarks in a proton changes as we change $\muf$:
\begin{equation}
\label{eq:quarkevolution}
\muf\frac{d}{d\muf}\, f_{a/A}(\xi',\muf) = 
\frac{\as(\muf)}{\pi}\int_{\xi'}^1\!\frac{d\xi}{\xi}\,
f_{a/A}(\xi,\muf)\,P_{aa}^{(1)}(\xi'/\xi)
\;.
\end{equation}
where
\begin{equation}
P_{aa}^{(1)}(\xi'/\xi) = G(\xi'/\xi,\bm 0)
\end{equation}
and we have set the scale of $\as$ to $\muf$.

If we consider other perturbative processes and other kinds of parton splittings, $\bar q \to \bar q + \Lg$, $\Lg \to \Lg + \Lg$ and $\Lg \to q + \bar q$, we obtain an evolution equation of the form
\begin{equation}
\label{eq:partonevolution}
\muf\frac{d}{d\muf}\, f_{a/A}(\xi',\muf) = 
\frac{\as(\muf)}{\pi}\sum_b \int_{\xi'}^1\!\frac{d\xi}{\xi}\,
f_{b/A}(\xi,\muf)\,P_{ab}^{(1)}(\xi'/\xi)
+ \cO(\as^2)
\;.
\end{equation}
As we have seen, this evolution equation for the parton distribution functions is needed if the factorization of PDFs from hard scattering cross sections in Sec.~\ref{sec:DISPDFs} is to be consistent with perturbative QCD. The equation is known as the Dokshitzer-Gribov-Lipatov-Altarelli-Parisi (DGLAP) equation \cite{GribovLipatov, AltarelliParisi, Dokshitzer}. 

The DGLAP equation can be generalized to higher perturbative orders:
\begin{equation}
\label{eq:partonevolutionfull}
\muf\frac{d}{d\muf}\, f_{a/A}(\xi',\muf) = 
\sum_b \int_{\xi'}^1\!\frac{d\xi}{\xi}\,
f_{b/A}(\xi,\muf)\,P_{ab}(\as(\muf),\xi'/\xi)
\;.
\end{equation}
The function $P_{ab}(\as(\muf),\xi'/\xi)$ has an expansion in powers of $\as$:
\begin{equation}
P_{ab}(\as(\muf),\xi'/\xi) = \sum_{n=1}^\infty
\left(\frac{\as(\muf)}{\pi}\right)^n
P_{ab}^{(n)}(\xi'/\xi)
\;.
\end{equation}
%

\section{Parton distribution functions}
\label{sec:PDFdefinition}

The parton distribution functions can be defined using the field operators for quarks and gluons \cite{PDFdef}. For quarks of flavor $a$ as partons in a proton, the definition is
\begin{equation}
\label{eq:fapdef}
f_{a/\Lp}(\xi,\muf) =
\frac{1}{2} \int \frac{dy^-}{2\pi}\ e^{-\mi \xi P^+ y^-}
\frac{1}{2}\sum_s
\langle P,s| {\bar\psi}_a(0,y^-,{\bf 0}) 
\gamma ^+ 
F
{\psi}_a(0)|P,s\rangle
\;.
\end{equation}
Here $P$ is the proton momentum in a reference frame in which $P = (P^+, m_\Lp^2/(2 P^+), \bm 0)$. Then $|P,s\rangle$ is the state vector for the proton with spin $s = \pm 1/2$. We sum over $s$ and divide by 2, thus considering an unpolarized proton. We let $\psi_a(x)$ be the field operator for quarks of flavor $a$. (The field operator $\psi_a(x)$ also has a color index and a Dirac index, which we do not indicate explicitly.) The operator $\psi_a(x)$ destroys a quark in the proton. Then  ${\bar\psi}_a(0,y^-,{\bf 0})$ recreates this quark. By taking a Fourier transform with the factor $\exp(-\mi \xi P^+ y^-)$, we require that the quark had plus-momentum $\xi P^+$. The operator ${\bar\psi}_a(y)$ is evaluated at $y^+ = \bm y = 0$, which is equivalent to integrating over the minus and transverse components of the quark momentum. The definition also includes a Dirac gamma matrix $\gamma^+$. If we simply set the factor $F$ to 1, the factors in this definition are arranged so that in one way of making a quantum field theory for QCD, $f_{a/\Lp}(\xi,\mu_F)$ is the probability density for finding a quark of flavor $a$ carrying a momentum fraction $\xi$ in the proton.

However, the definition with $F = 1$ is not invariant under gauge transformations as given in Eq.~(\ref{eq:gaugetransformation}). To make the definition gauge invariant, we define
\begin{equation}
F =
{\cal P}\exp\!\left(
\mi g\int_0^{y^-}\! dz^- \sum_{a=1}^8{A}_a^+(0,z^-,{\bf 0})\, t_a
\right)
.
\end{equation}
Here $A_a^\mu(z)$ with $a = 1,\dots,8$ is the QCD vector potential, which creates and destroys gluons. The ${\cal P}$ indicates that the field operators are path ordered: the exponential is to be expanded and then field operators with $z^-$ closer to $y^-$ are moved to the left of field operators with $z^-$ closer to $0$.

One way of understanding the factor $F$ is to write it as
\begin{equation}
F =
\bar{\cal P}\exp\!\left(
-\mi g\int_{y^-}^\infty\! dz^- \sum_{a=1}^8{A}_a^+(0,z^-,{\bf 0})\, t_a
\right)
{\cal P}\exp\!\left(
\mi g\int_0^{\infty}\! dz^- \sum_{a=1}^8{A}_a^+(0,z^-,{\bf 0})\, t_a
\right)
.
\end{equation}
Here $\cal P$ orders larger values of $z^-$ to the left and $\bar{\cal P}$ orders smaller values of $z^-$ to the left. With this form, we can think of the quark in the proton as not being annihilated but rather as being scattered, so that it moves along a lightlike path to $z^- = \infty$. The quark carries color, so that it can absorb or emit gluons as it travels along this path. The interactions of the scattered quark are idealized, so that when it absorbs and emits gluons it is not deflected from its path. Then the idealized quark returns to $(0,y^-, \bm 0)$ to join the proton. With this viewpoint, the parton distribution function for a quark is very similar to the scattering of a quark in deeply inelastic scattering.

For antiquarks and for gluons, there are analogous definitions of $f_{a/\Lp}(\xi,\muf)$ that make use of different combinations of the field operators of QCD \cite{PDFdef}.

The points at which the field operators in Eq.~(\ref{eq:fapdef}) and in the definitions for antiquarks and gluons are evaluated are lightlike separated. This has the result that many of the Feynman diagrams that contribute to $f_{a/\Lp}(\xi,\muf)$ have ultraviolet divergences that are not connected simply to the running of $\as$ or simple field strength renormalizations. To eliminate these divergences, one needs to renormalize the products of field operators. One can do this using the $\MSbar$ renormalization prescription. This defines $\MSbar$ parton distribution functions. Effectively, this puts an upper limit on the transverse momentum of emitted partons in Eq.~(\ref{eq:fapdef}), $|\bm \ell| \lesssim \muf$. This renormalization prescription gives a precise meaning to Eq.~(\ref{eq:PDFchange}) that works at any order of perturbation theory.

The $\MSbar$ renormalization prescription for $f_{a/\Lp}(\xi,\muf)$ gives a differential equation that tells how $f_{a/\Lp}(\xi,\muf)$ changes when $\muf$ changes. This is the DGLAP equation (\ref{eq:partonevolutionfull}).

One cannot use the definition (\ref{eq:fapdef}) and the analogous definitions for antiquarks and gluons to calculate parton distribution functions in perturbation theory. The problem is that the momentum transfers involved in binding quarks and gluons into a proton are small, so that the corresponding values of the running $\as$ are of order 1, rendering perturbation theory not useful. One can, however, use these definitions to calculate parton distribution functions in lattice gauge theory. This is a very difficult calculation. One finds results of roughly the right size, but with large errors.

We do need accurate numerical values for the parton distribution functions. For this, we can consider the PDFs $f_{a/\Lp}(\xi,\mu_0)$ at a rather low scale $\mu_0$ to be unknown functions that need to be fit. Given $f_{a/\Lp}(\xi,\mu_0)$, we know $f_{a/\Lp}(\xi,\muf)$ at any higher scale by using the DGLAP evolution equation. The functions $f_{a/\Lp}(\xi,\muf)$ at higher scales are related to experimental results using Eqs.~(\ref{eq:DISfactorization}) and (\ref{eq:ABfactorization}). Now we can use the experimental results to fit the PDFs at the starting scale $\mu_0$.

\section{Jets}
\label{sec:jets}

In Sec.~\ref{sec:Splittings}, we saw that the singularity structure of Feynman diagrams for QCD leads to the expectation that the final state in high momentum transfer events will contain jets, collections of particles that are almost collinear with each other. Such jets are indeed seen. They are very clear in event displays from the Large Hadron Collider (LHC).

One can define cross sections to make jets in, for instance, p-p collisions at the LHC. Here we use a reference frame in which the momenta of the momenta $p_\LA$ and $p_\LB$ of the beam hadrons lie along the + and $-$ $z$ axes, with $\vec p_\LB = - \vec p_\LA$. Letting $p$ be the momentum of a jet, we let $p_\LT$ be the absolute value of its transverse momentum, let $y = (1/2)\log(p^+/p^-)$ be its rapidity, and let $\phi$ is its azimuthal angle around the $z$-axis. We define differences in azimuthal angles, $\phi_i - \phi_j$, by adding or subtracting $2\pi$ so that $-\pi < \phi_i - \phi_j < \pi$.
Then with a suitable jet definition, one can define the cross section $d\sigma/(d p_\LT\, d\phi\, dy)$ to make one jet plus anything. One can also define other cross sections involving jets.

For this, one needs a precise definition of what a jet is. Several definitions are available. A commonly used definition for hadron-hadron collisions is the {\em anti-$k_t$} algorithm \cite{antikt}. To define jets using this algorithm, we first pick a parameter $R$. A possible choice is $R = 0.7$.  Then we start with a list of {\em protojets}, which can be individual hadrons or else, in a perturbative calculation, individual partons. Each protojet $i$ has a momentum $p_i$.  The protojets consist of all hadrons with momenta in an allowed range in the detector or else all scattered partons with momenta in the allowed range in the perturbative final state. We also start with a list of jets, with each jet $k$ having momentum $p_k$. Initially the list of jets is empty. Now we follow a simple algorithm:
\begin{enumerate}
\item For each pair $i,j$ of protojets, define
\begin{equation}
d_{ij} = \min\!\left(\frac{1}{p_{i,\scT}^2}, \frac{1}{p_{j,\scT}^2}\right)
\frac{(y_i - y_j)^2 + (\phi_i - \phi_j)^2}{R^2}
\;.
\end{equation}
For each protojet $i$, define 
\begin{equation}
d_i = \frac{1}{p_{i,\scT}^2}
\;.
\end{equation}

\item Find the smallest of the $d_{ij}$ and the $d_i$. Call it $d_\mathrm{min}$.

\item If $d_\mathrm{min}$ is one of the $d_{ij}$, merge protojets $i$ and $j$ into a new protojet $k$ with
\begin{equation}
p_k = p_i + p_j
\;.
\end{equation}

\item If $d_\mathrm{min}$ is one of the $d_i$, the protojet $i$ is {\em not mergable}. Remove it from the list of protojets and add it to the list of jets.

\item If protojets remain, to 1.

\end{enumerate}
Since the number of protojets decreases by 1 at each step, eventually there are no more protojets and we have a list of jets. Many of the jets defined this way will have very small transverse momenta. These jets are not of interest. One is interested a jet with a large $p_\scT$, or perhaps two jets with large $p_\scT$, or perhaps all jets with $p_\scT$ larger than some specified minimum value. Then we can use this algorithm to measure, or calculate, the cross section to find a specified jet configuration plus any other hadrons.

Note that when two protojets are almost collinear with each other, they are immediately combined to a single protojet that has momentum equal to the sum of the two protojet momenta. When a protojet $i$ has a momentum with very small $p_{i,\scT}$, its momentum may contribute to the momentum of a high $p_\scT$ jet that has a similar $y$ and $\phi$ to the protojet, but, since $p_{i,\scT}$ is very small, the effect of this contribution on the jet momentum is negligible. For these reasons, a jet cross section defined with this jet algorithm is infrared safe in the sense of Sec.~\ref{sec:InfraredSafety} and Eq.~(\ref{eq:IRsafetyAB}).

\section{Factorization}
\label{sec:factorization}

We now return to Eqs.~(\ref{eq:Afactorization}) and (\ref{eq:ABfactorization}) in Sec.~\ref{sec:DISPDFs}, expressing PDF factorization in lepton-hadron collisions and hadron-hadron collisions.

The functions $f_{a/A}(\xi_A, \muf)$ are the functions defined as the proton matrix element of certain QCD operators in Sec.~\ref{sec:PDFdefinition}. In this way, the PDFs are defined independently of the particular cross section in which they appear, so that we do not have to be concerned that different PDFs might be needed for different cross sections. The parton cross sections $\hat\sigma$ can be expanded in powers of $\as$. For hadron-hadron collisions, 
\begin{equation}
\hat\sigma_{ab}[F;\xi_A, \xi_B,\muf]
= \sum_{n = 0}^\infty 
\left[\frac{\as(\mur)}{\pi}\right]^{n_\scB + n}
\hat\sigma_{ab}^{(n_\scB + n)}[F;\xi_A, \xi_B,\muf,\mur]
\;.
\end{equation}
Here $n_\scB$ is the power of $\as$ in the lowest order, or Born level, cross section. For instance, $n_\scB = 0$ for the Drell-Yan cross section but $n_\scB = 2$ for the cross section to produce two jets. The statement of PDF factorization asserts that the functions $\hat\sigma_{ab}^{(n_\scB + n)}[F;\xi_A, \xi_B,\muf,\mur]$ are calculable using QCD perturbation theory and have neither ultraviolet nor infrared singularities.

It should be noted that the calculation of $\hat\sigma_{ab}^{(n_\scB + n)}[F;\xi_A, \xi_B,\muf,\mur]$ beyond the leading order $\as^{n_\scB}$ is not simple. At intermediate stages of the calculation in $4 - 2\epsilon$ dimensions, there are ultraviolet poles $1/\epsilon^N$. These are removed by $\MSbar$ renormalization. There are also infrared poles $1/\epsilon^N$ that are removed once we account for the presence of the parton distribution functions.

These formulas must work at any order of perturbation theory, even if $\hat \sigma$ is only calculated at a rather low order. This is because if the factorization formulas were to fail beyond, say, $\as^{n_\scB + n}$, then we would expect to find contributions proportional to $\as^{n_\scB + n + 1}$ that reflect physics at a low scale of order $m_\Lp$. Then if the hard interaction has scale $Q$, this contribution could be proportional to $\as(Q)^{n_\scB}$ times $\as(m_\Lp)^{n+1}$, where $\as(m_\Lp)$ is not small but is of order $\as(m_\Lp) \sim 1$. If this happened, we could not trust the low order prediction.

It is reasonably straightforward to argue that the DIS factorization formula (\ref{eq:Afactorization}) works. Here there is only one hadron in the initial state. However, it is much more difficult to verify the factorization formula (\ref{eq:ABfactorization}) for two hadrons in the initial state. 

\begin{figure}[t]
\begin{center}
\includegraphics[width = 10 cm]{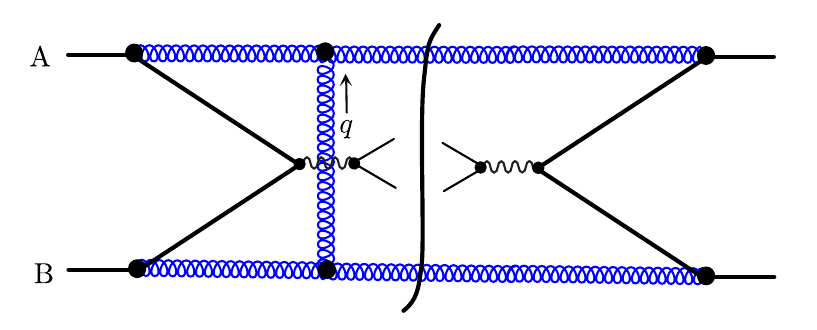}
\end{center}
\caption{
Graph for $A + B \to \mu^+ + \mu^-$ with a Glauber singularity. This illustrates the quantum amplitude and the conjugate amplitude, which create the cross section. The slightly curved vertical line represents the final state. Only one initial state parton from each hadron is shown.}
\label{fig:spectatorspectator}
\end{figure}

Why could there be a problem? Consider the Feynman diagram in Fig.~\ref{fig:spectatorspectator} for the Drell-Yan process, $A + B \to \mu^+ + \mu^- + X$. Hadron $A$ has very large momentum in the +direction, while hadron $B$ has very large momentum in the $-$direction. Long before the hard interaction that produces the muon pair, a quark or antiquark in hadron $A$ has emitted a gluon with large momentum close to the direction of hadron $A$. In addition, long before the hard interaction, an antiquark or quark in hadron $B$ has emitted a gluon with large momentum close to the direction of hadron $B$. The two gluons are just spectators and do not participate directly in the hard interaction. The spectator gluons exchange a gluon with momentum $q$. One needs to integrate over $q^+$, $q^-$, and $\bm q$. A simple analysis shows that this graph makes a leading contribution to the Drell-Yan cross section rather than a contribution suppressed by a power of $m_\Lp/Q$. Furthermore, the graph does not have the factorized form of Eq.~(\ref{eq:ABfactorization}). In a gauge theory, there are identities known as Ward identities that relate different graphs and can often yield cancellations among different graphs. However, these identities are useful when $q^+$ and $q^-$ are of the same rough size as $|\bm q|$. In this graph, the leading contribution comes from what is called the Glauber integration region, $q^+ \ll |\bm q|$ and $q^- \ll |\bm q|$. 

\begin{figure}[t]
\begin{center}
\includegraphics[width = 4.5 cm]{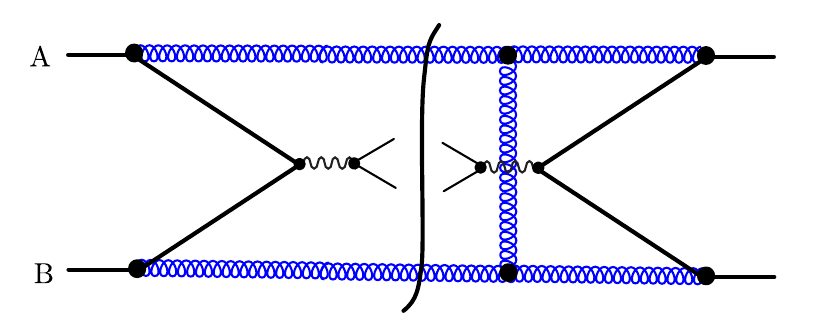}
\includegraphics[width = 4.5 cm]{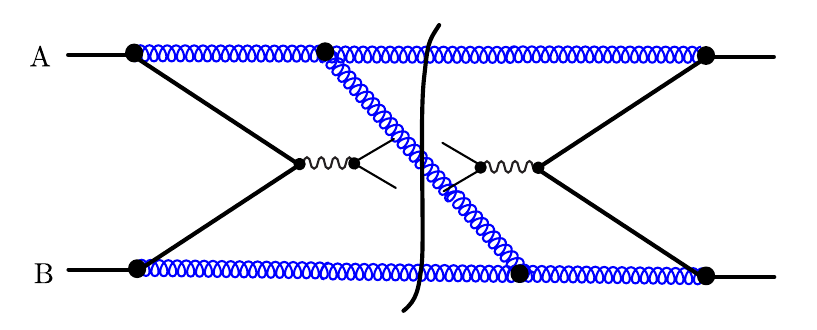}
\includegraphics[width = 4.5 cm]{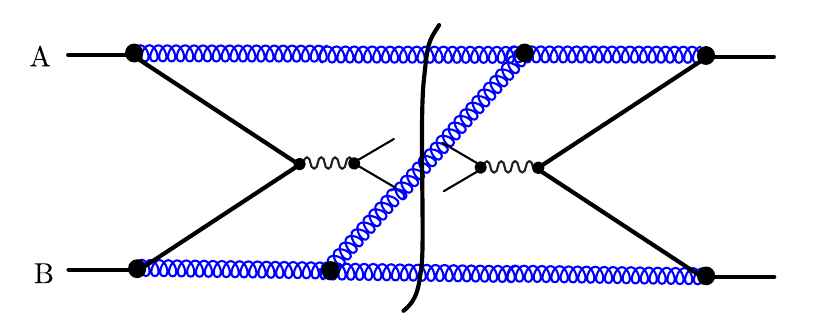}
\end{center}
\caption{
Graphs for $A + B \to \mu^+ + \mu^-$ to be combined with the graph from Fig.~\ref{fig:spectatorspectator}.}
\label{fig:spectatorgraphs}
\end{figure}

One key step needed to address this problem is to combine the graph of Fig.~\ref{fig:spectatorspectator} with other graphs shown in Fig.~\ref{fig:spectatorgraphs}, in which the gluon is exchanged in the conjugate Feynman amplitude, the gluon emitted by hadron A into the final state in the Feynman amplitude and is emitted by hadron B into the final state in the conjugate Feynman amplitude, or else the gluon is emitted with the roles of A and B reversed. It is important here that the observable is infrared safe, so that the appearance of a gluon with a very small momentum $\pm q$ in the final state does not affect the measurement.

Detailed arguments supporting Eq.~(\ref{eq:ABfactorization}) were presented in Refs.~\cite{Bodwin1985, CSS1985, CSS1988}. These arguments cannot be said to constitute a rigorous mathematical proof, but they do address the problems in some detail.




\begin{thebibliography}{99}
\bibitem{ZweigQuarks}
G.~Zweig,
``An SU(3) model for strong interaction symmetry and its breaking. 
Version 2,''
doi:10.17181/CERN-TH-412.
\end{thebibliography}
\end{document}